\documentclass{aa}
\usepackage{graphicx}
\usepackage[svgnames]{xcolor}
\usepackage[pdftex,colorlinks,citecolor=DarkBlue]{hyperref}
\usepackage{txfonts}
\usepackage{mathrsfs}
\usepackage{amsmath}
\usepackage{amssymb}
\usepackage{orcidlink}
\usepackage[switch]{lineno}
\usepackage{soul}

\usepackage{makecell}
\usepackage{subfigure}
\usepackage{diagbox}
\bibpunct{(}{)}{;}{a}{}{,}
	
\begin{document} 
   \title{A new wideband radio polarization observation of the Supernova Remnant G315.4$-$2.3}

   \titlerunning{wideband polarization observation of G315.4$-$2.3}

   \author{
           X. Chen\inst{\ref{ynu}}\and
           X. Sun\inst{\ref{ynu}}\and
           J. F. Kaczmarek\inst{\ref{skao}}\and
           B. M. Gaensler\inst{\ref{uc}}$^,$\inst{\ref{di}}$^,$\inst{\ref{daa}}
           \and
           P. Slane\inst{\ref{cfa}} \and
           J. L. West\inst{\ref{drao}}
    }
   \institute{School of Physics and Astronomy, Yunnan University, Kunming 650500, PR China\\\email{chenxin\_@mail.ynu.edu.cn, xhsun@ynu.edu.cn}\label{ynu}
    \and
   SKA Observatory, SKA-Low Science Operations Centre, 26 Dick Perry Avenue, Kensington, WA 6151, Australia\label{skao}
   \and
   Department of Astronomy and Astrophysics, University of California, Santa Cruz, 1156 High Street, Santa Cruz, CA 95064, USA\label{uc}
   \and
   Dunlap Institute for Astronomy and Astrophysics, University of Toronto, 50 St. George Street, Toronto, ON M5S 3H4, Canada\label{di}
    \and
    David A.\ Dunlap Department of Astronomy and Astrophysics, University of Toronto, 50 St. George Street, Toronto, ON M5S 3H4, Canada\label{daa}
    \and
   Center for Astrophysics | Harvard \& Smithsonian, 60 Garden Street, Cambridge, MA 02138, USA\label{cfa}
   \and
   Dominion Radio Astrophysical Observatory, Herzberg Astronomy \& Astrophysics, National Research Council Canada, P.O. Box 248, Penticton, BC V2A 6J9, Canada\label{drao}
    }

\date{Accepted XXX. Received YYY; in original form ZZZ}
 
  \abstract
    {The supernova remnant (SNR) G315.4$-$2.3 (MSH~14$-$63 or RCW~86) exhibits strong emission across the electromagnetic spectrum. Radio polarization observations probe magnetic fields and will help to understand the evolution of the SNR.}
   {We aim to investigate the radio spectrum and magnetic field properties of the SNR.}
    {We observed G315.4$-$2.3 using the Australia Telescope Compact Array (ATCA), covering the frequency range of 1.1-3.1~GHz. We then performed rotation measure (RM) synthesis on the frequency cubes of $Q$ and $U$ to obtain the polarized intensity and RM. The regular component of the line-of-sight magnetic field was estimated from RM. The fractional polarization versus wavelength squared was used to constrain the properties of the turbulent magnetic field.}
   {We obtained image cubes of Stokes $I$, $Q$, and $U$ over the frequency range 1.319-3.023~GHz after excluding channels affected by radio frequency interference, images of polarized intensity $P$ and RM, and images of fractional polarization $p$. The rms noise is 0.6~mJy~beam$^{-1}$ for the band-averaged $I$ and is 90~$\mu$Jy~beam$^{-1}$ for $P$. All images have been smoothed to a common resolution of $62\arcsec\times33\arcsec$. The comparison with single-dish observations indicates that our images have retained the larger-scale emission. We obtained a spectral index of $\alpha=-0.60\pm0.03$ for the SNR. The radio spectra are very similar for different areas of the SNR. The foreground RM was estimated to be approximately 55~rad~m$^{-2}$, and the internal RM of most SNR areas is small, less than about 50~rad~m$^{-2}$. The regular magnetic field along the line of sight was estimated to be about 1.4~$\mu$G in the southwest, much smaller than the total magnetic field. For most parts of the southwest and northeast, $p$ is less than 8\% and is nearly constant with $\lambda^2$. We estimated the ratio of turbulent to regular magnetic field to be larger than about 3. The scale of the turbulent magnetic field for some area in the northwest might be smaller than about 0.4~pc.}
   {The radio characteristics, including spectrum and turbulent magnetic field, are very similar in the northeast and southwest, even though the evolution is quite different for these two regions based on the current models. These should be taken into account for future modeling of the evolution of the SNR.}

   
   \maketitle

\section{Introduction}
The object G315.4$-$2.3 was first identified as a supernova remnant (SNR) by \citet{1967+Hill} due to its nonthermal and polarized emission. It is also referred to as MSH~14$-$63 as it was cataloged by \citet{1961+Mills} or RCW~86 \citep{1960+Rodgers}, which is the bright optical nebula located in the southwest region of the source.

G315.4$-$2.3 is a very interesting SNR, showing strong emission over almost the full electromagnetic spectrum, from TeV $\gamma$-ray~\citep{2018+Abramowski}, GeV $\gamma$-ray~\citep{2014+Yuan}, X-ray~\citep{bamba+2023, suzuki+22, 2017+Tsubone, borkowski+2001}, optical~\citep{1997+Smith}, infrared~\citep{2011+Williams}, to radio~\citep{2001+Dickel} bands. The origin of the TeV $\gamma$-ray emission remains uncertain, and hadronic~\citep{2019+Sano} and leptonic~\citep{2016+Ajello} scenarios have been proposed. X-ray emission contains thermal and nonthermal components toward the northeast, northwest, and southwest parts of the SNR~\citep{2002+Rho, 2006+Vink, 2008+Yamaguchi, 2013+Castro, 2017+Tsubone}. The nonthermal X-ray emission requires electrons of TeV energy which can only be accelerated with a high shock velocity. Toward the northeast, the shock velocity was measured to be 3000~km~s$^{-1}$~\citep{Yamaguchi+16} to 6000~km~s$^{-1}$~\citep{2009+Helder}. However, the shock velocity towards the southwest is much smaller and only in the range of 300-2000~km~s$^{-1}$~\citep{suzuki+22}. The way in which the electrons are accelerated there with such a low shock velocity is puzzling. What is even more puzzling is the circular shape of the SNR from radio to X-ray bands, given that the shock velocities differ significantly between the northeast and southwest.     

G315.4$-$2.3 is also an interesting target due to its connection with the SN AD~185 in Chinese ancient records~\citep{1975+Clark}, as such connections are very rare. This puts its age to be 1840~yr. Observations such as nonradiative H$\alpha$ filaments~\citep{1997+Smith}, Fe-rich ejecta~\citep{2002+Rho, 2007+Ueno, 2008+Yamaguchi, 2014+Broersen}, and the nondetection of any associated neutron star~\citep{kaplan+04} indicate that G315.4$-$2.3 is a remnant of \uppercase\expandafter{\romannumeral1}a SN. \citet{2011+Williams} and \citet{2014+Broersen} conducted hydrodynamic simulations showing that a \uppercase\expandafter{\romannumeral1}a SN inside a cavity formed by the progenitor system can reproduce the multi-wavelength SNR morphology and the high-energy $\gamma$-ray emission. In this scenario, due to either an off-center explosion or an inhomogeneous interstellar medium density, the northeast part encountered the cavity wall only recently and the southwest started to interact with the wall about 300~yr after the explosion. This scenario provides an explanation for the difference in the shock velocities in the two regions. Alternatively, \citet{2017+Gvaramadze} claimed the identification of a solar-type star in a binary system with a companion neutron star and thus suggested that G315.4$-$2.3 is from a core-collapse SN. However, earlier observations by \citet{migani+2012} had ruled out the companion as a neutron star. Therefore, the \uppercase\expandafter{\romannumeral1}a SN scenario is more favored than the core-collapse SN scenario.

Radio polarization observations provide information on the spectrum and magnetic field, which helps to understand the emission and evolution of the SNR. However, there have only been a few polarization observations of G315.4$-$2.3 to date. Earlier observations were conducted with the Parkes 64-m telescope. \citet{milne+1975} observed the SNR at 5~GHz with a resolution of $4\farcm4$ and detected strong polarized emission towards the east and southwest regions. The fractional polarization was about 5\% and the magnetic field was approximately in the radial direction. \citet{1976+Dickel} observed the SNR at 2.7~GHz with a resolution of $8\farcm4$ and found that the fractional polarization is very similar to that at 5~GHz towards the east and southwest regions. 

Observations with an interferometer provide a higher-resolution image of the SNR. The SNR was observed by \citet{2001+Dickel} with the Australian Telescope Compact Array (ATCA) at 1.34~GHz with a resolution of $8\arcsec$, and polarized emission was detected at a few spots toward the east and southwest regions. Based on the six 32-MHz bands, an average Faraday rotation measure (RM) of about 60~rad~m$^{-2}$ was obtained. At a similar resolution, but with a much broader bandwidth covering the frequency range of 856-1712~MHz, \citet{2024+Cotton} observed the SNR with MeerKAT. They found very similar features in total intensity, but with much higher sensitivity and dynamic range than that observed by \citet{2001+Dickel}. They also detected more polarized emission and thus obtained RMs for almost the entire SNR shell. The RM values are consistent with those of \citet{2001+Dickel}.

In this paper, we report a new wideband polarization observation of G315.4$-$2.3 covering the frequency range of 1.1--3.1~GHz, conducted with the ATCA, which bridges the frequency gap of the observations currently available. The paper is organized as follows: observations and data reduction are presented in Sect.~\ref{sec:observations}; results are shown in Sect.~\ref{sec:results}; discussions are made in Sect.~\ref{sec:discussions}; conclusions are drawn in Sect.~\ref{sec:conclusions}.

\section{Observations and data reduction}
\label{sec:observations} 

We observed G315.4$-$2.3 with the ATCA~(Project ID: C3246) using the Compact Array Broad-band Backend~\citep[CABB,][]{wilson+11}. The observations covered the frequency range 1.1--3.1~GHz split into 2048 1-MHz channels. Since G315.4$-$2.3 has an angular size much larger than the primary beam, the mosaicking mode was used for the observations. In total, 43 pointings were observed to achieve full spatial Nyquist sampling up to 3.1~GHz. To achieve good $uv$ coverage, six array configurations were used: H75, H168, EW352, 750C, 1.5A and 1.5D, with baselines ranging from 30.6~m to about 1.5~km. The observations were conducted seven times from April to September 2018. The total integration time was $\sim$0.4~h per pointing.

\begin{figure}
\centering
\includegraphics[width=0.9\columnwidth]{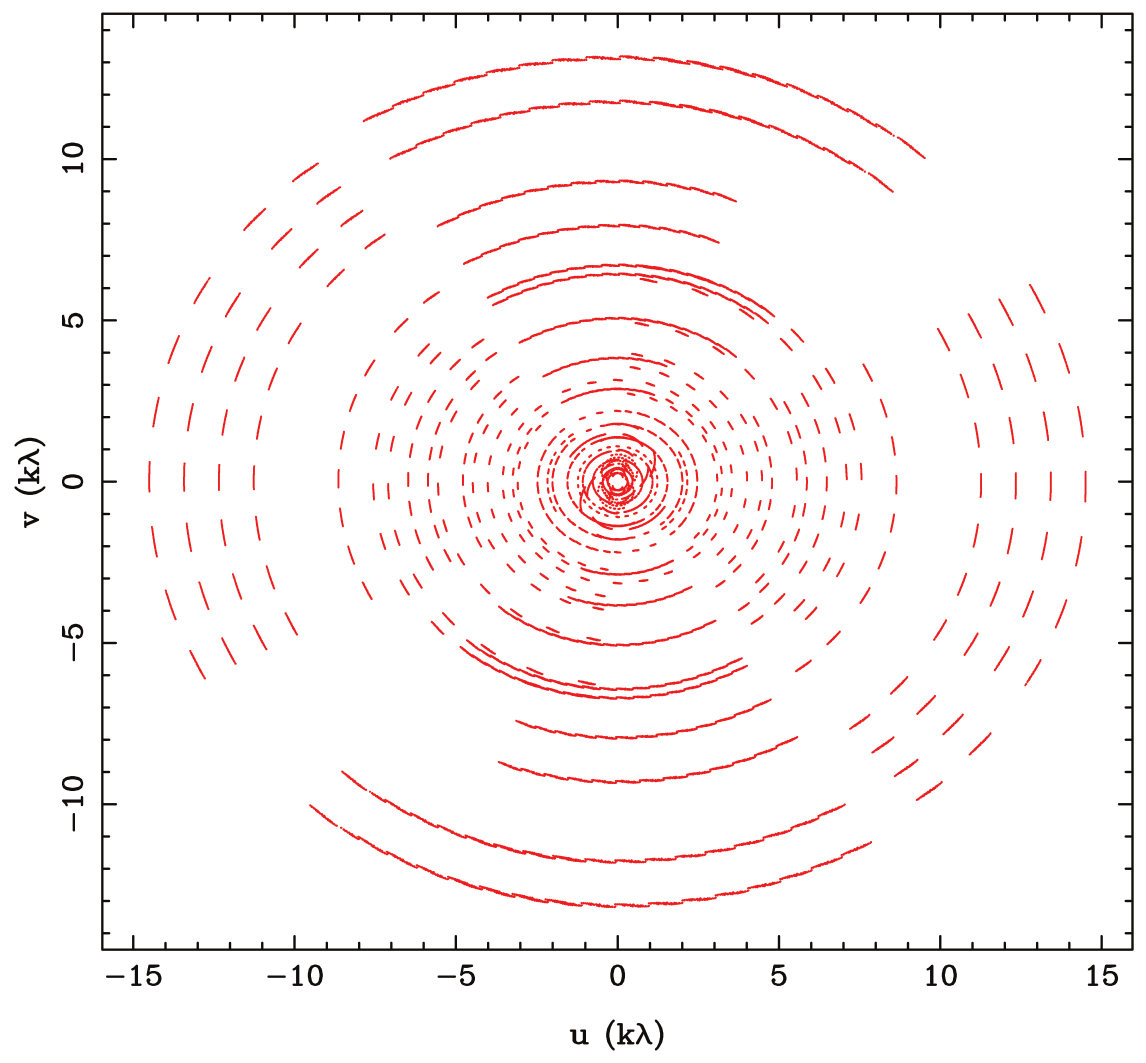}
    \caption{$uv$ coverage of the 16 frequency channels centered at 3.023~GHz from all the pointings of G315.4$-$2.3.}
    \label{fig: UV coverage}
\end{figure}

Calibration and imaging were performed using \texttt{MIRIAD}~\citep{1995+Sault}. Bandpass solutions were derived from 1934$-$638 or 0823$-$500. Gain and polarization leakage were solved for using the phase calibrator 1352$-$63, with observations spanning a wide range of parallactic angles. The phase calibrator was observed approximately every 16~minutes and gain solutions were derived over the same time interval. Observations of the targets were bracketed by observations of the phase calibrator, and the phase solution was interpreted between observations of the calibrator. To account for the frequency-dependent variation of the gains, we split the 2~GHz bandwidth into eight subbands to derive the solutions separately. These solutions were scaled to match the flux density model of 1934$-$638 and then applied to the targets before imaging.

 Our observations were also influenced by radio frequency interference (RFI), with strong RFI always present, especially at the low-frequency end of the band. After flagging, approximately 40\% of the data was discarded and all data at 1.1--1.3~GHz were flagged. 

We imaged Stokes $I$, $Q$, and $U$ by joint deconvolution of all the pointings every 16 consecutive channels, using the \texttt{MIRIAD} tasks \texttt{MOSMEM} and \texttt{PMOSMEM}. The Briggs weighting~\citep{Briggs+1995} with $\rm robust=-1$ was used. The $uv$-coverage for the 16 channels centered at the high frequency end of 3.023~GHz is shown in Fig.~\ref{fig: UV coverage}, where the minimum $uv$ distance is about 0.2~k$\lambda$, corresponding to a maximum angular scale of approximately $17\arcmin$. 

We obtained 80 images of $I$, $Q$, and $U$ using 16-MHz bandwidth each, ranging from 1.319~GHz to 3.023~GHz. We smoothed the resultant images to a common resolution of $62\arcsec\times33\arcsec$. For $I$, we averaged all the channels to obtain the image centered at 2.2~GHz. For $Q$ and $U$, we applied RM synthesis~\citep{2005+Brentjens} to obtain the polarized intensity and RM. 

According to \cite{Burn+66}, the following Fourier transform relation holds: $\mathcal{P}(\lambda^2)=Q(\lambda^2)+iU(\lambda^2)=\int_{-\infty}^\infty F(\phi)\exp(2i\phi\lambda^2){\rm d}\phi$, where $\mathcal{P}$ is the observed complex polarized intensity and its absolute value is the polarized intensity; $\phi(\vec{r})=K\int_{\vec{r}}^{\rm observer}n_eB_\parallel{\rm d}l$ is the Faraday depth, $K$ is a constant, $n_e$ is the thermal electron density, $B_\parallel$ is the line-of-sight component of the magnetic field, and the integral is along the line of sight from a position $\vec{r}$ in the source to the observer; $F(\phi)$ is the Faraday spectrum with 
the absolute value $|F(\phi)|$ representing polarized intensity as a function of Faraday depth $\phi$. For polarized emission, $|F(\phi)|$ peaks at the RM of the emission with the peak value corresponding to the polarized intensity ($P$). 

RM synthesis derives $F(\phi)$ from the observed $\mathcal{P}(\lambda^2)$ using the Fourier transform. Due to the limited bandwidth, $\mathcal{P}$ is not fully sampled and correspondingly there is an RM spread function (RMSF) in the Faraday depth domain. The full width at half maximum (FWHM) of the RMSF contributes to the uncertainty of RM. Details on RM synthesis and calculation of the FWHM of the RMSF refer to \cite{2005+Brentjens}.

We used RM-Tools~\citep{Purcell+2020} to perform RM synthesis and then search for the peaks in $|F(\phi)|$ to derive RM and $P$. The intrinsic polarization angle ($\chi_0$) after correction for the Faraday rotation with the derived RM was also obtained. 

The bandwidth of the previous ATCA observations of G315.4$-$2.3 by \citet{2001+Dickel} is too narrow, and the frequency band has been covered by our new observations. The MeerKAT observations missed large-scale emission as shown later in Sect.~\ref{sec:spec}. We thus did not include these two observations to perform RM synthesis and fractional polarization analysis.

\section{Results}
\label{sec:results}

\subsection{Total intensity and spectral index}
\label{sec:spec}
\begin{figure}[!htbp]
    \includegraphics[width=\columnwidth]{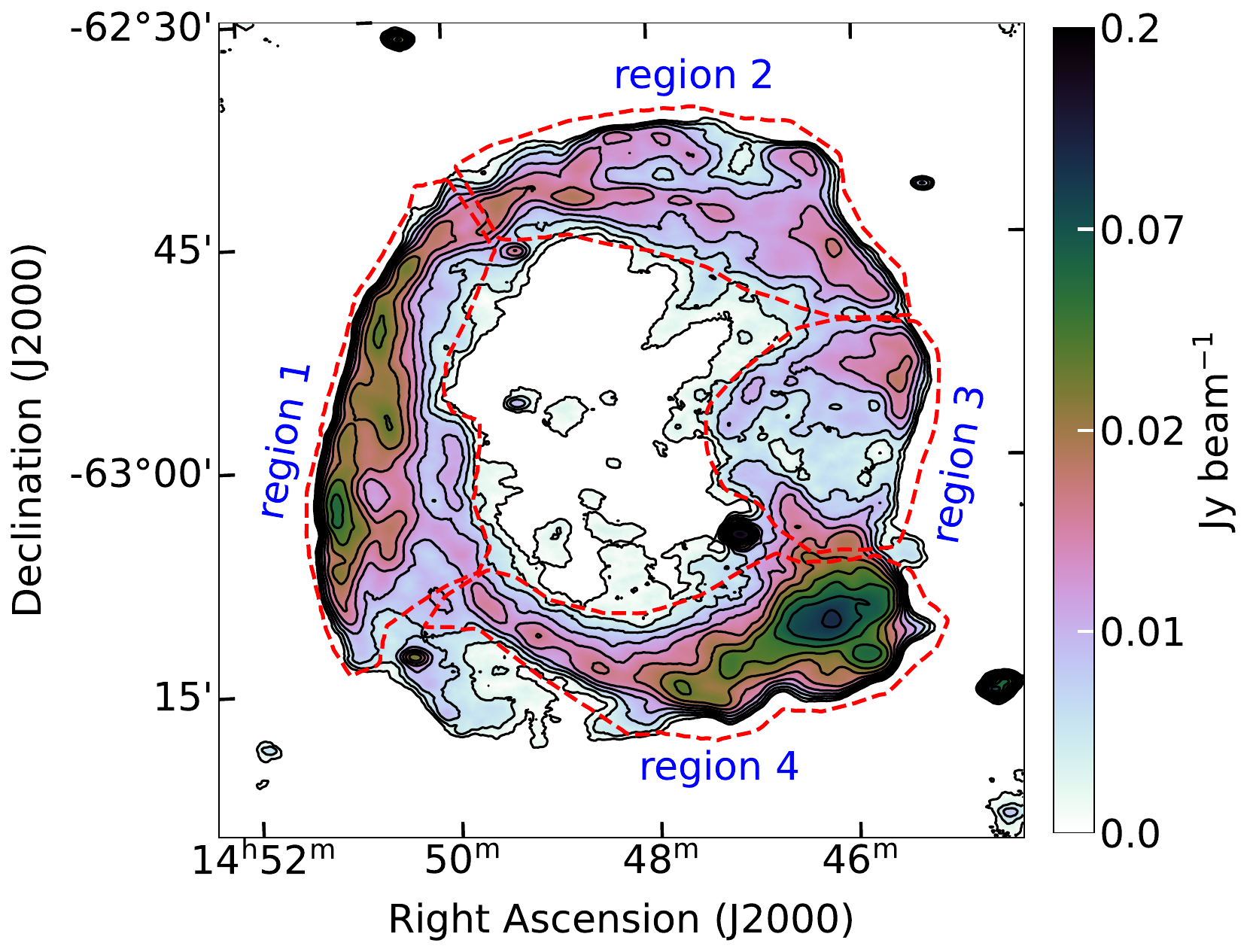}
    \caption{Total intensity ($I$) image and contours of G315.4$-$2.3 at 2.2~GHz from averaging all the frequency channels. The contour levels are at $2^{0.5n}\times5\sigma_I$, $n=0,1,2,\ldots$, and $\sigma_I=0.6$~mJy~beam$^{-1}$ is the rms noise. The resolution is $62^{\arcsec}\times33^{\arcsec}$. The red dashed lines mark the four regions, for which the spectral indices were derived and shown in Fig.~\ref{fig:g315_4regions_SpecIndex}.}
    \label{fig:total intensity}
\end{figure}

The total intensity ($I$) image of G315.4$-$2.3 at 2.2~GHz is shown in Fig.~\ref{fig:total intensity}, which has a resolution of $62\arcsec\times33\arcsec$ with a position angle (PA) of $84\degr$. The rms noise is $\sigma_I=0.6$~mJy~beam$^{-1}$. It shows an obvious circular shell-like structure with the two brightest segments toward the northeast and southwest, similar to the radio images of \cite{2001+Dickel} and \cite{2024+Cotton}. It also resembles the X-ray~\citep{2017+Tsubone,2006+Vink,borkowski+2001} and optical~\citep{1997+Smith} images.

\begin{table}[!htbp]
\centering
\renewcommand{\arraystretch}{1.2}
\caption{Integrated flux density of SNR~G315.4$-$2.3.}
\label{tab:G315_flux}
\small
\begin{tabular}{lccc}
\hline\hline
 $\nu$ (MHz) & $S_\nu$ (Jy)  & Telescope & Reference \\
\hline
 408 & 86$\pm$9\tablefootmark{a} & MOST &1\\
843 & 22.7$\pm$1.4\tablefootmark{b} & MOST & 2\\
1335 & 15.4$\pm$1.0  & MeerKAT &3 \\
1340 & 28$\pm$3\tablefootmark{a}  & ATCA &4\\
1410 & 30$\pm$3\tablefootmark{a}  & Murriyang & 5 \\
2200 & 21.7$\pm$1.1 &ATCA & 6 \\
2400 & 24.3$\pm$1.0\tablefootmark{b}  & Murriyang & 7\\
2650 & 20$\pm$2\tablefootmark{a}  & Murriyang & 5\\
4850 & 10.1$\pm$0.5\tablefootmark{b}  & Murriyang & 8 \\
5000 & 18.2$\pm$1.8\tablefootmark{a}  &Murriyang &1\\
\hline
\end{tabular}
\tablebib{(1)~\citet{1975+Caswell}; (2) \citet{1996+Whiteoak}; (3) \citet{2024+Cotton}; (4) \citet{2001+Dickel}; (5) \citet{1967+Hill}; (6) this paper; (7) \citet{1995+Duncan}; (8) \citet{1993+Condon};}
\tablefoot{
MOST: Molonglo Observatory Synthesis Telescope; Murriyang: the Parkes 64-m telescope
\tablefoottext{a}{The uncertainty was not provided in the literature and we used a relative uncertainty of 10\%.}
\tablefoottext{b}{We measured the integrated flux density from the survey maps.}
}
\end{table}

\begin{figure}[!htbp]	
    \includegraphics[width=\columnwidth]{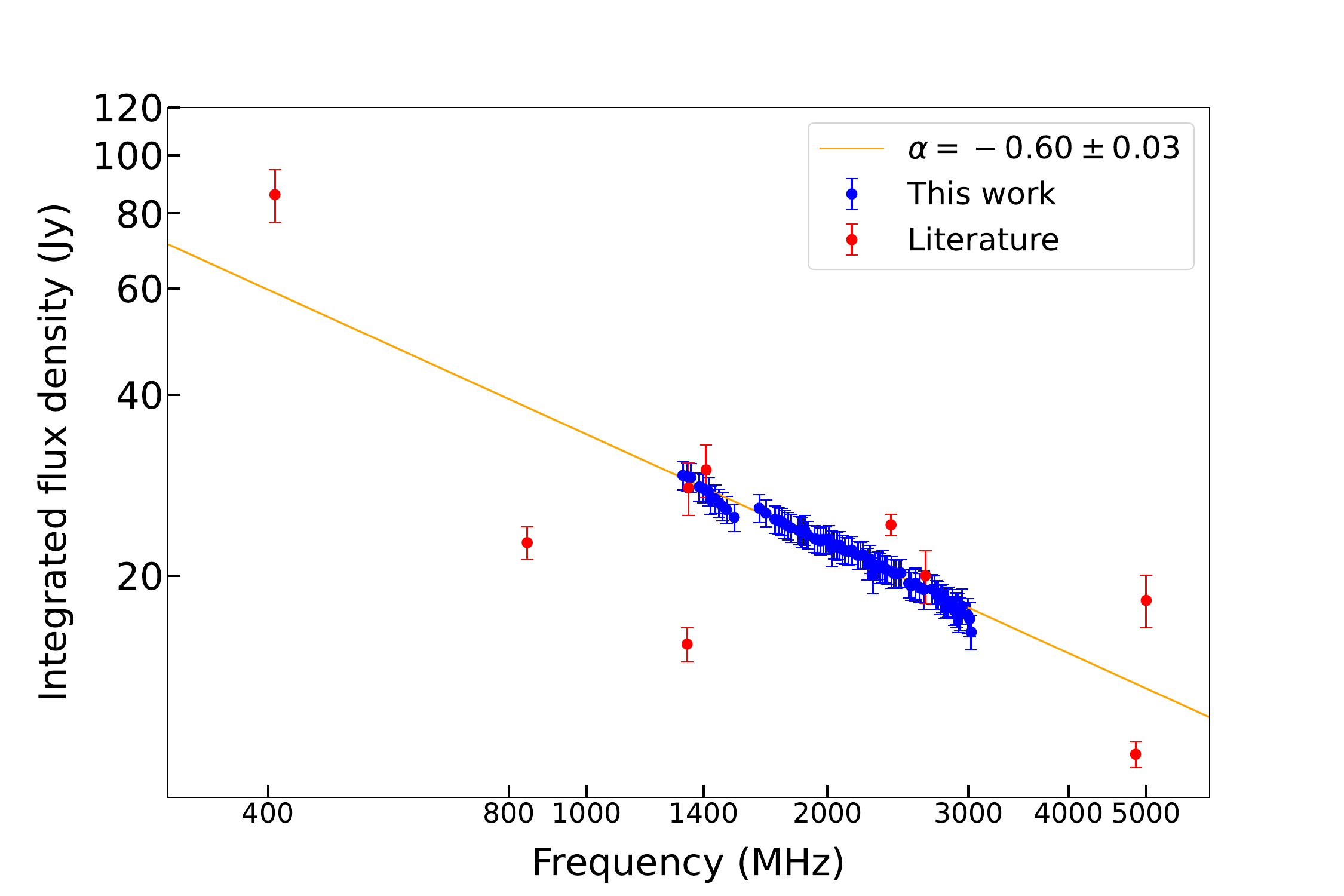}
    \caption{Integrated flux density versus frequency for SNR~G315.4$-$2.3. Blue points were measured from the 80 $I$ images and red points were from Table~\ref{tab:G315_flux}.
    }
    \label{fig:g315_SpecIndex}
\end{figure}

We measured an integrated flux density of 21.7$\pm$1.1~Jy at 2.2~GHz from Fig.~\ref{fig:total intensity}. We also collected previously published flux density values or measured flux density from the available surveys. The integrated flux density $S_\nu$ and the frequency $\nu$ are listed in Table~\ref{tab:G315_flux}. We also measured the integrated flux density from all the 80 ATCA images. The uncertainties of the measurements include a conservative 5\% calibration error~\citep{2018+Butler}. All the flux densities ($S_\nu$) versus frequencies ($\nu$) are shown in Fig.~\ref{fig:g315_SpecIndex}.

\begin{figure*}	[!htbp]
    \centering  \includegraphics[width=0.9\textwidth]{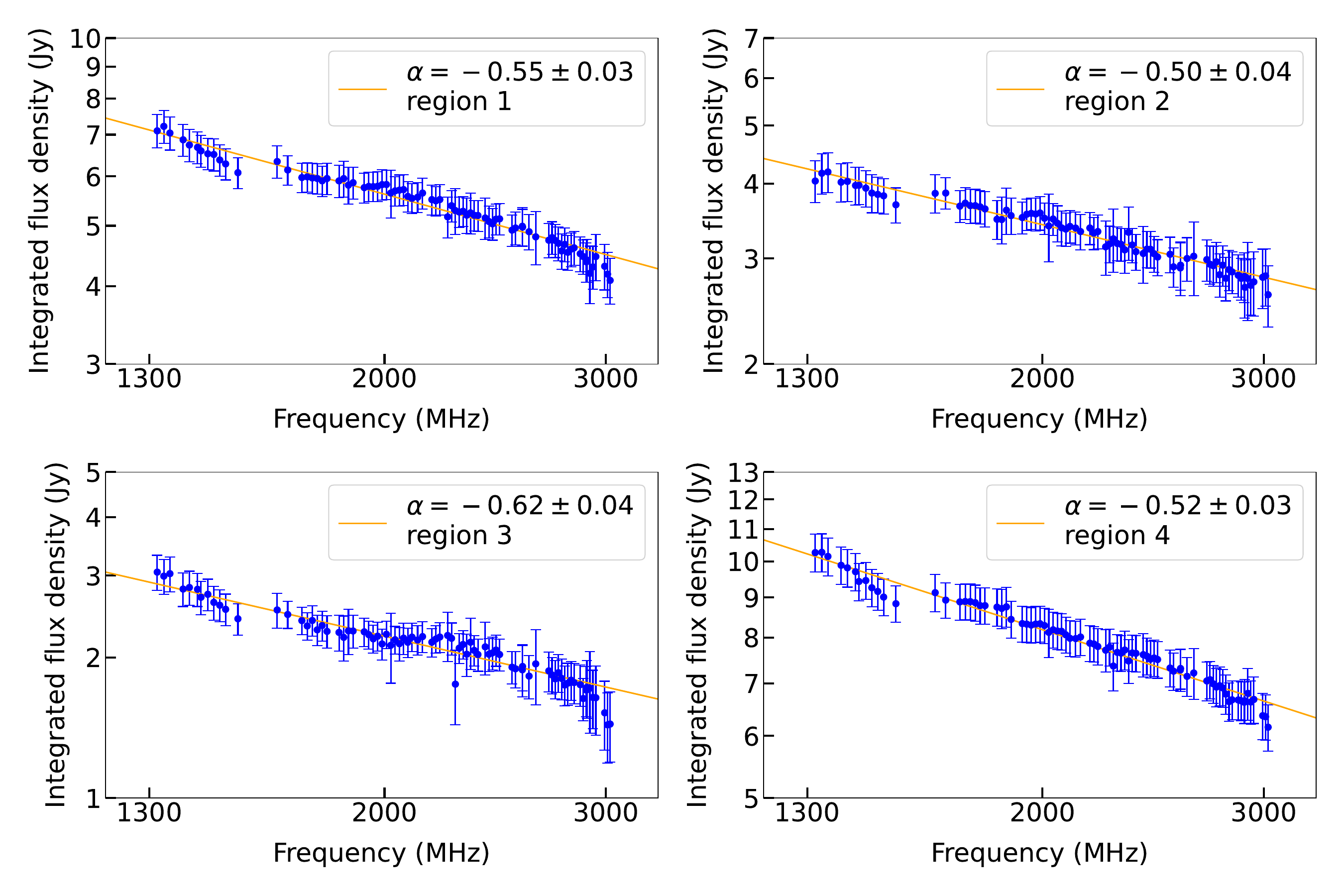}
    \caption{The same as Fig.~\ref{fig:g315_SpecIndex}, but for the four regions marked in Fig.~\ref{fig:total intensity} and showing only the new measurements from ATCA.}
 \label{fig:g315_4regions_SpecIndex}
\end{figure*}

The new measurements of integrated flux density are consistent with those from single-dish observations by Murriyang at 1.41, 2.4, and 2.65~GHz, indicating that the missing flux density with ATCA is negligible across the entire band. This is the result expected from the joint deconvolution that delivers a maximum angular scale of $\lambda/(d_{\rm min}-D/2)$, where $d_{\rm min}$ is the minimum baseline and $D=22$~m is the diameter of the ATCA telescope~\citep[e.g.][]{McClure-Griffiths+01}. Even at the high frequency end of 3.023~GHz with $d_{\rm min}=0.2$~k$\lambda$~(Fig.~\ref{fig: UV coverage}), the maximum angular scale is about $38\arcmin$, roughly the size of G315.4$-$2.3. Therefore, all the flux can be recovered.

We fit the integrated flux density versus frequency to a power law, $S_\nu\propto\nu^\alpha$, and obtained the spectral index  $\alpha=-0.60\pm0.03$. This is consistent with $\alpha=-0.62$ obtained by \citet{1975+Caswell} based on the measurements at 85.5, 1410, and 2650~MHz, although their values are larger than our fit shown in Fig.~\ref{fig:g315_SpecIndex}. 

Several measurements deviate from the fitting as can be seen in Fig.~\ref{fig:g315_SpecIndex}. The integrated flux densities at 408~MHz and 5000~MHz~\citep{1975+Caswell} are larger than the fitting, but have large uncertainties~(Table~\ref{tab:G315_flux}), and the maps were too small in angular size to accurately subtract background emission. The three measurements at 843~MHz by MOST, at 1335~MHz by MeerKAT, and at 4850~MHz by Murriyang all miss flux density and fall below the fitted line. For MOST, there was a lack of baselines shorter than 15~m,  which made the observations insensitive to structures on scales larger than $\sim30\arcmin$~\citep{1996+Whiteoak}. For the MeerKAT observation, the imaging used an inner Gaussian taper, which reduced sensitivity to large-scale emission~\citep{2024+Cotton}. For the Murriyang observation, the data processing employed a $57\arcmin$ running-median baseline subtraction, which suppressed structures with angular scales larger than $\sim30\arcmin$~\citep{1993+Condon}.

To investigate spatial variations, the remnant was divided into four regions outlined in Fig.~\ref{fig:total intensity}. The integrated flux densities were derived for each individual region and for all the 80 channels. The spectral index was also obtained by fitting the data to a power law. The results are shown in Fig.~\ref{fig:g315_4regions_SpecIndex}. Taking into account the uncertainties, the difference among the spectral indices is small. 

\subsection{Polarized intensity and RM}

\begin{figure}	
    \includegraphics[width=\columnwidth]{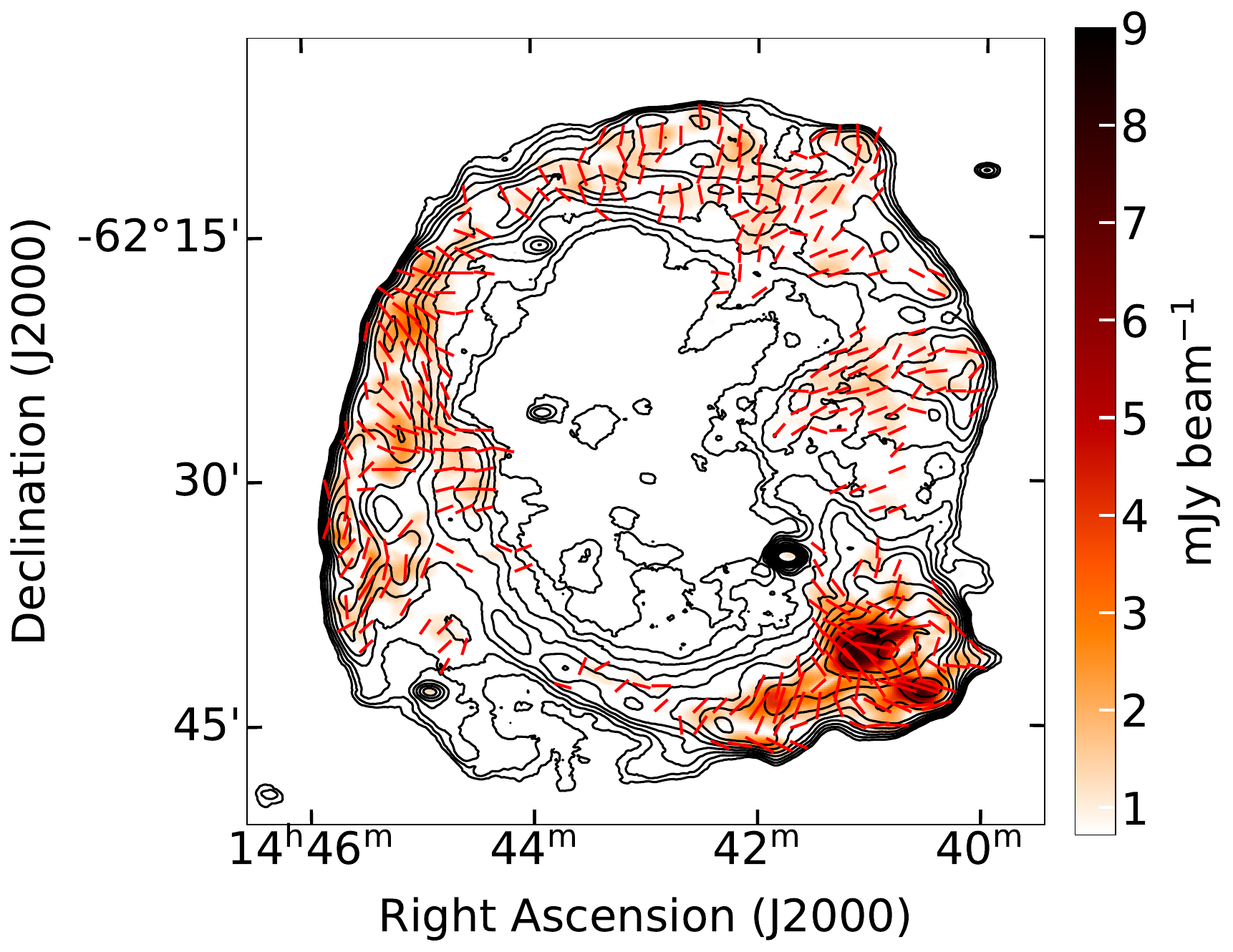}
    \caption{Image of $|F(\phi_{\rm peak})|$, corresponding to polarized intensity $P$, superimposed with $I$ contours of G315.4$-$2.3. The contour levels are the same as Fig.~\ref{fig:total intensity}. The bars indicate the orientation of magnetic fields with lengths proportional to $P$.}
    \label{fig:Peak polarization}
\end{figure}

\begin{figure}
	\includegraphics[width=\columnwidth]{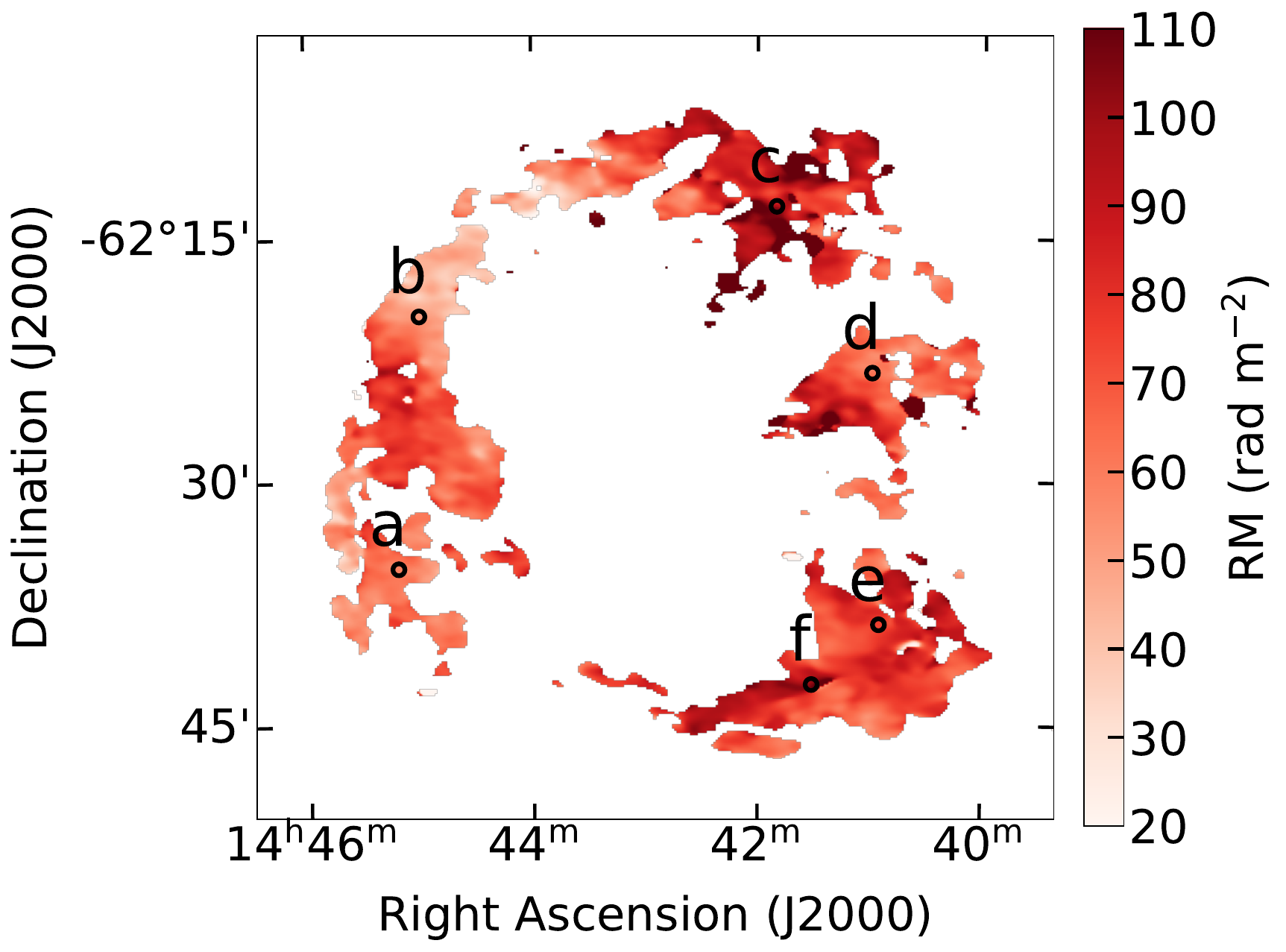}
    \caption{RM ($\phi_{\rm peak}$) image of G315.4$-$2.3. For the pixels marked with a-f, the Faraday spectra were extracted from the cube and shown in Fig.~\ref{fig:check RM}.}
    \label{fig:RM}
\end{figure}

We obtained a cube of Faraday depth from $-$500 to $+$500~rad~m$^{-2}$ in steps of 5~rad~m$^{-2}$, and images of $\phi_{\rm peak}$, $|F(\phi_{\rm peak})|$, and intrinsic polarization angle $\chi_0$ from RM synthesis. For $|\phi|$ close to 500~rad~m$^{-2}$, the frames in the cube contain predominantly noise, and we measured from these frames an rms noise of $\sigma_P=90$~$\mu$Jy~beam$^{-1}$ for polarized intensity. We set a threshold of $|F(\phi_{\rm peak})|>8\sigma_P$, which means the RM uncertainty is better than ${\rm FWHM}/2{\rm SNR}_P\approx5$~rad~m$^{-2}$. Here, ${\rm FWHM}\approx83$~rad~m$^{-2}$ determined by the maximum separation of $\lambda^2$ of the 80 frequency channels~\citep{2005+Brentjens}, and ${\rm SNR}_P$ is the signal-to-noise ratio for $P$. 

We show $|F(\phi_{\rm peak})|$ in Fig.~\ref{fig:Peak polarization} and $\phi_{\rm peak}$ in Fig.~\ref{fig:RM}. The orientation of the magnetic field ($\chi_0+90\degr$) is also depicted in Fig.~\ref{fig:Peak polarization}. The Faraday spectra for pixels from different parts of the SNR covering a wide range of $\phi_{\rm peak}$ were extracted from the cube and shown in Fig.~\ref{fig:check RM}. These spectra exhibit simple profiles with a single dominant peak, and thus $\phi_{\rm peak}$ and $|F(\phi_{\rm peak})|$ correspond to RM and $P$, respectively. Note that with the large FWHM of the RMSF, simple profiles could hide the complexity that may be present. The complexity occurs when there is a mixture of synchrotron-emitting and Faraday-rotating medium or there are multiple emitting components, each with different Faraday depth~\citep[e.g.][]{Sun+15}. The complexity leads to wavelength-dependent depolarization~\citep{Sokoloff+98}, which was not observed, as shown later. This supports that the spectra are simple.    

We detected polarized emission from most of the shell of G315.4$-$2.3. The strongest polarized intensity appears in the southwest, where the magnetic field appears to be perpendicular to the shell. This was also observed by \citet{2001+Dickel} with ATCA and \citet{2024+Cotton} with MeerKAT. Besides the southwest, the previous ATCA observation only detected polarized emission from a few spots in the northeast~\citep{2001+Dickel}; the MeerKAT observation also detected polarized emission from most of the SNR revealing much finer details~\citep{2024+Cotton}.    

The RM shows a large variation across the SNR. Toward the southwest, the RM is around 80~rad~m$^{-2}$; toward the northeast, the RM decreases from about 70~rad~m$^{-2}$ in the lower part to about 45~rad~m$^{-2}$ in the upper part; toward the northwest, RM is the largest and around 90~rad~m$^{-2}$. The narrower band of the previous ATCA observation~\citep{2001+Dickel} does not allow for a reliable estimate of RM. The RM image from MeerKAT~\citep{2024+Cotton} is strikingly consistent with ours, but with finer details. This might imply that missing flux in total intensity does not influence the RM determination~\citep[e.g.][]{Ordog+25}.  

\subsection{Fractional polarization}

\begin{figure}
    \centering
    \includegraphics[width=0.95\columnwidth]{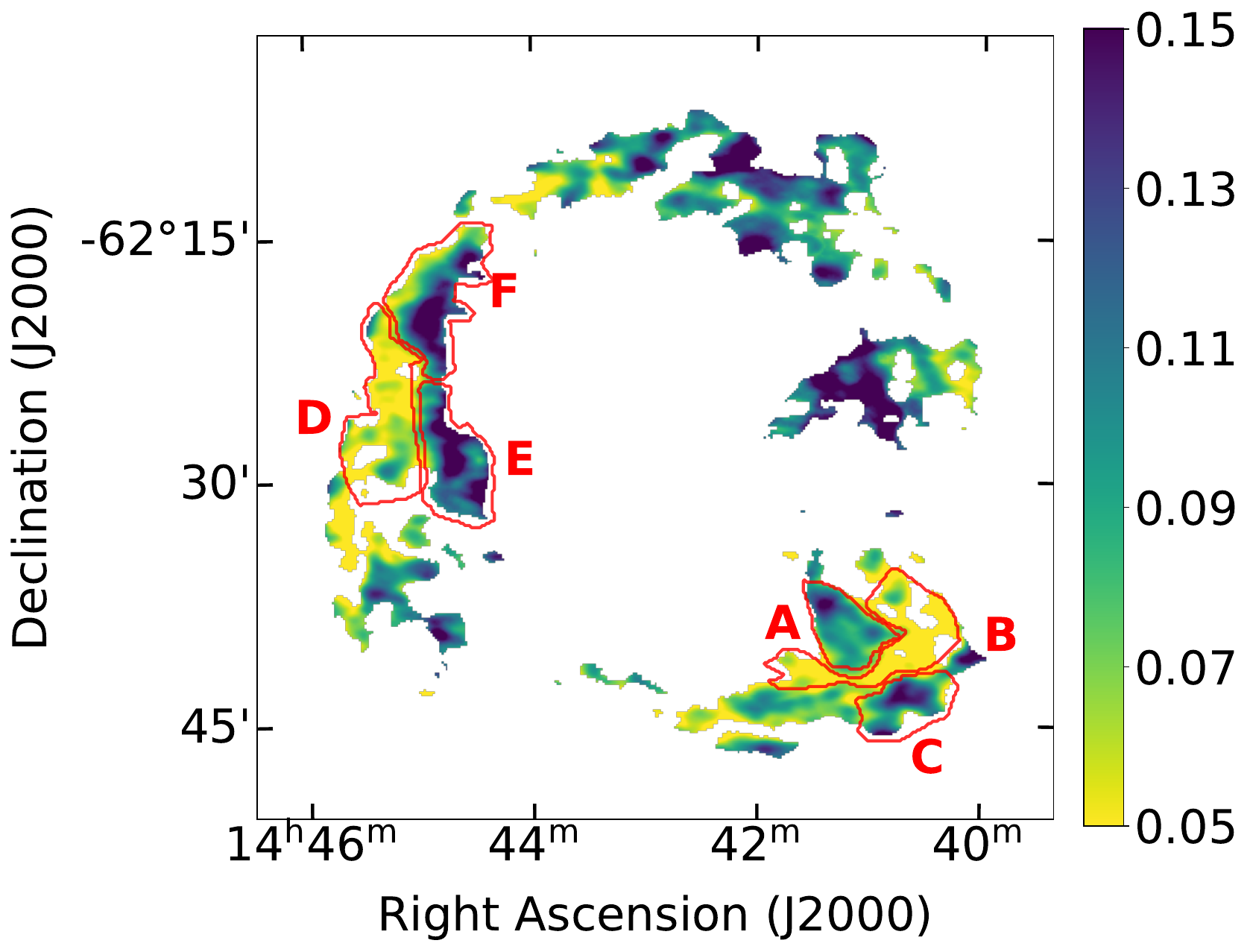}
    \caption{Fractional polarization ($p$) of G315.4$-$2.3 at 2.02~GHz. Several regions are outlined for analysis shown in Fig.~\ref{fig:p_lam2}.}    
    \label{fig:the_degree_of_polarization}
\end{figure}

The fractional polarization can be derived as $p=P/I$. Note that the polarized intensity from RM synthesis is at $\lambda_0$ which is the average of $\lambda^2$ of all channels~\citep{2005+Brentjens}. In our case, $\lambda_0$ corresponds to the frequency of 2.02~GHz. We scaled $I$ from 2.2~GHz~(Fig.~\ref{fig:total intensity}) to 2.02~GHz using the spectral index of $-0.6$~(Fig.~\ref{fig:g315_SpecIndex}). We also set thresholds of $I>5\sigma_I$, $P>8\sigma_P$, and $\sigma_p/p=\sqrt{(\sigma_P/P)^2+(\sigma_I/I)^2}<0.1$. The image of $p$ is shown in Fig.~\ref{fig:the_degree_of_polarization}.

Toward the northeast and southwest, there are mixed regions of low ($p\approx 3\%$) and high ($p\approx15\%$) fractional polarization; toward the northwest, the fractional polarization is mostly high at a level of larger than about 11\%. The median fractional polarization is $\sim 9\%$ for the entire SNR.

\section{Discussions}
\label{sec:discussions}

We adopted a distance $d=2.5$~kpc for G315.4$-$2.3, which is commonly referenced in literature~\citep[e.g.][]{2014+Broersen}. Published measurements give $d=2.3$~kpc from Balmer-dominated filament 
$V_{\rm LSR}$ kinematics \citep{2003+Sollerman} and $d=2.8$~kpc from optical/H\,I kinematics \citep{1996+Rosado}. We therefore adopt $d=2.5$~kpc as a representative midpoint.

\subsection{Foreground RM}

\begin{figure}	
    \centering
    \includegraphics[width=0.95\columnwidth]{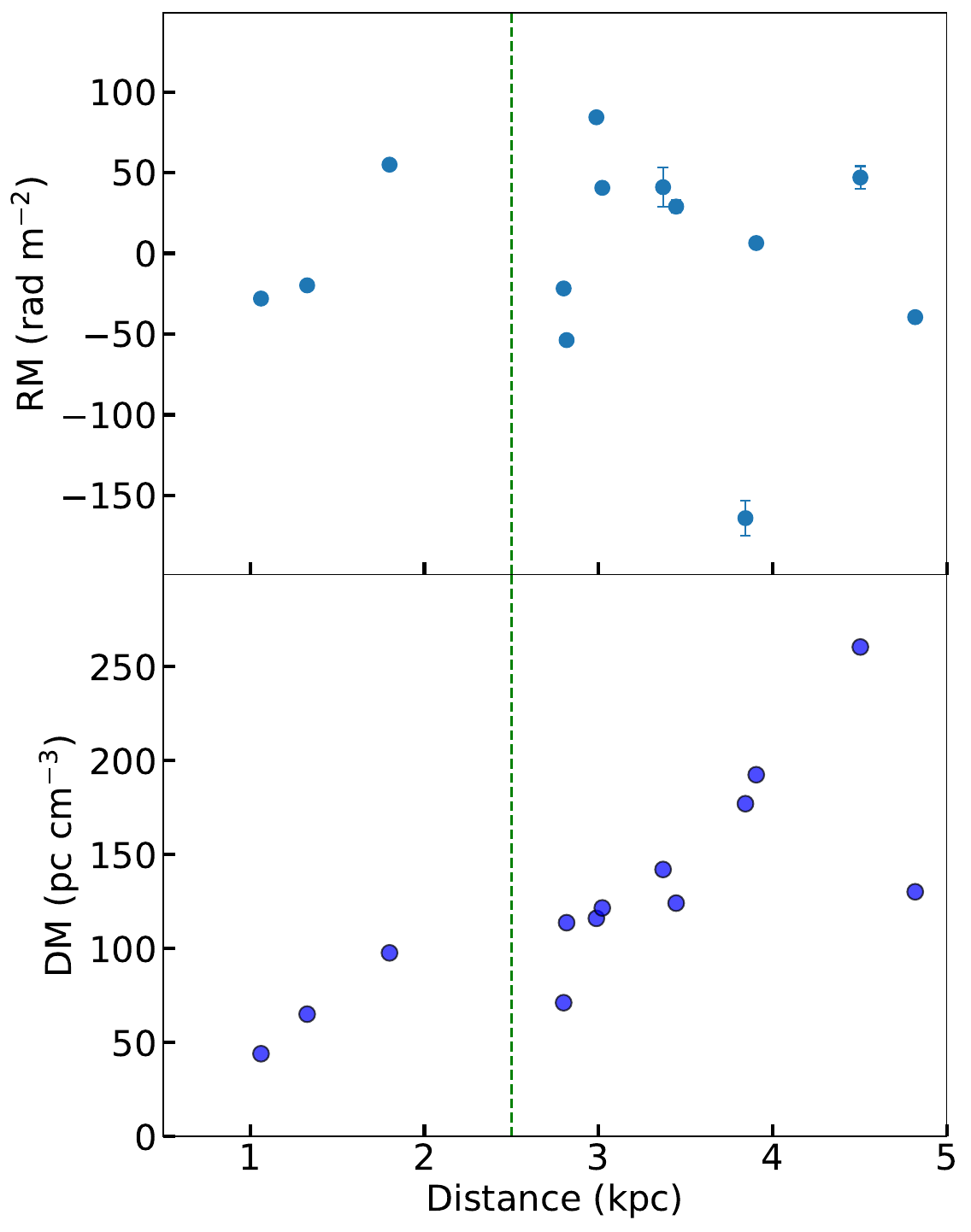}
    \caption{RM and DM versus distance for pulsars within $5\degr$ of G315.4$-$2.3 center. The green dashed line indicates the distance of 2.5 kpc.}
    \label{fig: pulsars_distribution}
\end{figure}

The RMs shown in Fig.~\ref{fig:RM} contain contributions from the Galactic foreground and the SNR itself, and it is necessary to separate them. We searched for pulsars from the ATNF pulsar catalog~\citep[][version 2.6.5]{2005+Manchester} within a $5\degr$ radius centered at G315.4$-$2.3 and plotted RM and dispersion measure (DM) versus distance in Fig.~\ref{fig: pulsars_distribution}. 

The RMs appear to increase until a distance of about 2~kpc reaching a value of about 55~rad~m$^{-2}$, and then show a large scatter beyond the distance of about 2.5~kpc; the DMs generally increase linearly, but the slope is nearly flat between 2 and 3~kpc. Consequently, we assume that the increase in RM between 2 and 3~kpc is also small, and we take the value of 55~rad~m$^{-2}$ as the foreground. Unfortunately, there are only 3 pulsars in front of the SNR, and the estimate of foreground RM is very uncertain. 

For a larger foreground RM, we would expect the average RMs of pulsars beyond 2~kpc to be systematically larger. Although this does not seem to be the case, as can be seen in Fig.~\ref{fig: pulsars_distribution}, this cannot be ruled out because of the small number of pulsars. 

For a much smaller foreground RM, the internal RM would be large, causing Faraday differential depolarization~\citep{Sokoloff+98}. However, the spatial pattern of the fractional polarization~(Fig~\ref{fig:the_degree_of_polarization}) does not follow that of the RM~(Fig.~\ref{fig:RM}), particularly for the northwest where the fractional polarization is higher, but the RM is also larger than the rest of the SNR. Provided the depolarization was primarily due to internal RM, the opposite would be expected since a larger RM results in a larger depolarization, and hence a lower fractional polarization. Therefore, a much smaller foreground RM is not favored.   

There is depolarization caused by the internal RM only in areas with large RMs around 100~rad~m$^{-2}$. The internal RM contributed by the SNR can reach about 50~rad~m$^{-2}$ towards the northwest after subtracting the foreground RM. Suppose a uniform mixture of thermal and relativistic electrons, the observed fractional polarization is expected to be $p_{\rm obs}=p_{\rm i}\sin(2{\rm RM}\lambda^2)/(2{\rm RM}\lambda^2)$, where $p_{\rm i}$ is the intrinsic fractional polarization~\citep{Sokoloff+98}. For $p_{\rm i}\approx70\%$ and $\rm RM=50$~rad~m$^{-2}$, $p_{\rm obs}$ is about 25\% at 2.02~GHz, larger than the observed values toward the northwest~(Fig.~\ref{fig:the_degree_of_polarization}), implying that there is further depolarization.  

\subsection{Magnetic field in the southwest}
\label{Sect:B_in_SW}

The magnetic field can be estimated on the basis of RM provided with the thermal electron density. The thermal electron density can be derived from the H$\alpha$ observations. We used the H$\alpha$ image compiled by \citet{2003+Finkbeiner}. There is only strong H$\alpha$ emission toward the southwest, where we can estimate the magnetic field. 

The emission measure (EM) can be derived as ${\rm EM}=2.75T_4^{0.9}I_{\rm H\alpha}\exp[2.44E(B-V)]$~\citep{1998+Haffner}, where $T_4$ is the electron temperature in $10^4$~K, $I_{\rm H\alpha}$ is the H$\alpha$ intensity in Rayleigh, $E(B-V)$ is the color excess, and EM is in pc~cm~$^{-6}$. Toward the southwest, we obtained ${\rm EM}\approx443$~pc~cm$^{-6}$ with background-subtracted $I_{\rm H\alpha}=54$~Rayleigh from the H$\alpha$ map \citep{2003+Finkbeiner}, $E(B-V)=0.53$ from the 3D extinction map~\citep{2019+Chen} and $T_4=0.8$~\citep{1985+Reynolds}. 

We assumed that the path length $L$ is the same as the width of the shell which is about $6\arcmin$ corresponding to $4.4$~pc with $d=2.5$~kpc. The electron density was then estimated to be $n_e=\sqrt{{\rm EM}/L}\approx10$~cm$^{-3}$ with a filling factor of 1, and the line-of-sight component of the magnetic field was estimated to be $B_\parallel\approx\mathcal{R}/0.81n_eL\approx1.4$~$\mu$G. Here, $\mathcal{R}=2{\rm RM}$ is the full RM and is twice the measured RM with thermal and nonthermal gas uniformly mixed~\citep{Sokoloff+98}, and $\rm RM\approx25$~rad~m$^{-2}$ after excluding the foreground contribution. The $B_\parallel$ derived this way represents only the regular component. 

The fitting of the spectral energy distribution of synchrotron emission including observations from radio, X-ray, and $\gamma$-ray by \citet{2016+Ajello} yielded a transverse magnetic field of $16.8\pm2.1$~$\mu$G for the southwest region. Toward the Galactic longitude of about $315\degr$ for a distance of 2.5~kpc, RM is mainly contributed by the Carina arm, and the line-of-sight component is approximately at the same level as the transverse component assuming that the magnetic field follows the spiral arm~\citep[][their Fig.~4]{Han+18}. This means that the total magnetic field is much larger than the regular field along line of sight. 

The regular $B_\parallel$ would become even smaller if the foreground RM were larger. In contrast, a smaller foreground RM would result in a larger regular $B_\parallel$. For example, a foreground RM of half the currently used value can increase the regular $B_\parallel$ by a factor 2, but a much smaller foreground RM introduces a larger depolarization that is not consistent with the observation. With a smaller filling factor of 0.1, the regular $B_\parallel$ increases by a factor $\sim$3. Combining these together, the regular $B_\parallel$ is still smaller than the total magnetic field, implying the presence of a strong turbulent magnetic field. 

Based on X-ray spectral analysis, \citet{suzuki+22} found that there exists highly amplified magnetic turbulence toward the southwest. This is consistent with our results.   

\subsection{Southwest versus northeast}

The shock velocity of the southwest~\citep{suzuki+22} is much smaller than that of the northeast~\citep{2009+Helder, Yamaguchi+16}, but the SNR retains a circular shape in radio and X-ray. The current models have suggested that the southwest interacted with a cavity wall shortly after the supernova explosion and thus decelerated, whereas the northeast just started to interact with the wall~\citep{2011+Williams, 2014+Broersen}. It is worthwhile to compare the radio characteristics of these two regions.  

\subsubsection{Spectrum} 

There is no large spatial variation of the spectral index in the SNR, ranging from $-0.5$ to $-0.6$. A leptonic model can fit the broadband emission from radio to TeV X-ray well and yielded the spectral index of electrons $\Gamma_e=2.3\pm0.1$~\citep{2018+Abramowski}, corresponding to $\alpha=-(\Gamma_e-1)/2=-0.65\pm0.05$. Fitting the broadband emission of the southwest and northeast separately with a leptonic model yielded the same $\Gamma_e=2.21\pm0.1$ for both regions~\citep{2016+Ajello}, corresponding to $\alpha=-0.61\pm0.05$. This value is slightly steeper than what we obtained: $\alpha=-0.55\pm0.03$ for the northeast (Fig.~\ref{fig:g315_4regions_SpecIndex}: region 1) and $\alpha=-0.52\pm0.03$ for the southwest (Fig.~\ref{fig:g315_4regions_SpecIndex}: region 4). The radio spectral indices for these two regions are also nearly the same.

Based on the current models, the southwest mainly contains an old population of electrons since the shock has long decelerated, and the northwest contains both old and young populations of electrons. Thus, it is expected that the spectrum of the southwest is steeper than that of the northeast because of synchrotron cooling. Spectral analysis of X-ray emission by \citet{2017+Tsubone} showed that the power-law spectral index of the southwest is larger than that of the northeast. Unfortunately, radio emission at GHz is typically from GeV electrons that have a large cooling time of the order of $10^6$~yr for a magnetic field of 20~$\mu$G as inferred from the fitting of broadband emission. Since the cooling time is much longer than the age of the SNR, the same radio spectra for the northeast and southwest support that the injected spectra of electrons accelerated by the SNR shocks are the same. 

\begin{figure}[!htbp]	
    \centering
    \includegraphics[width=0.95\columnwidth]{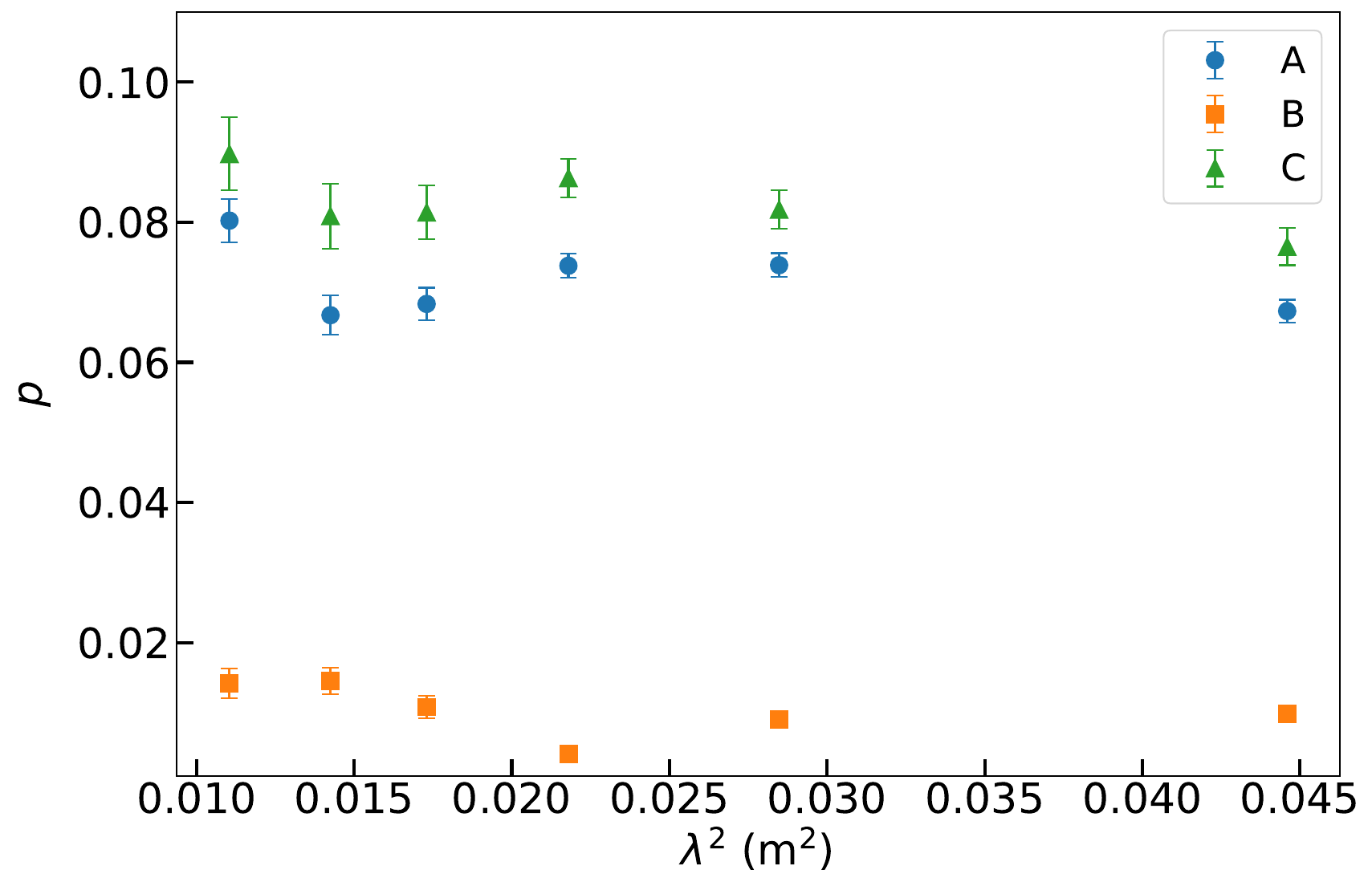}
    \includegraphics[width=0.95\columnwidth]{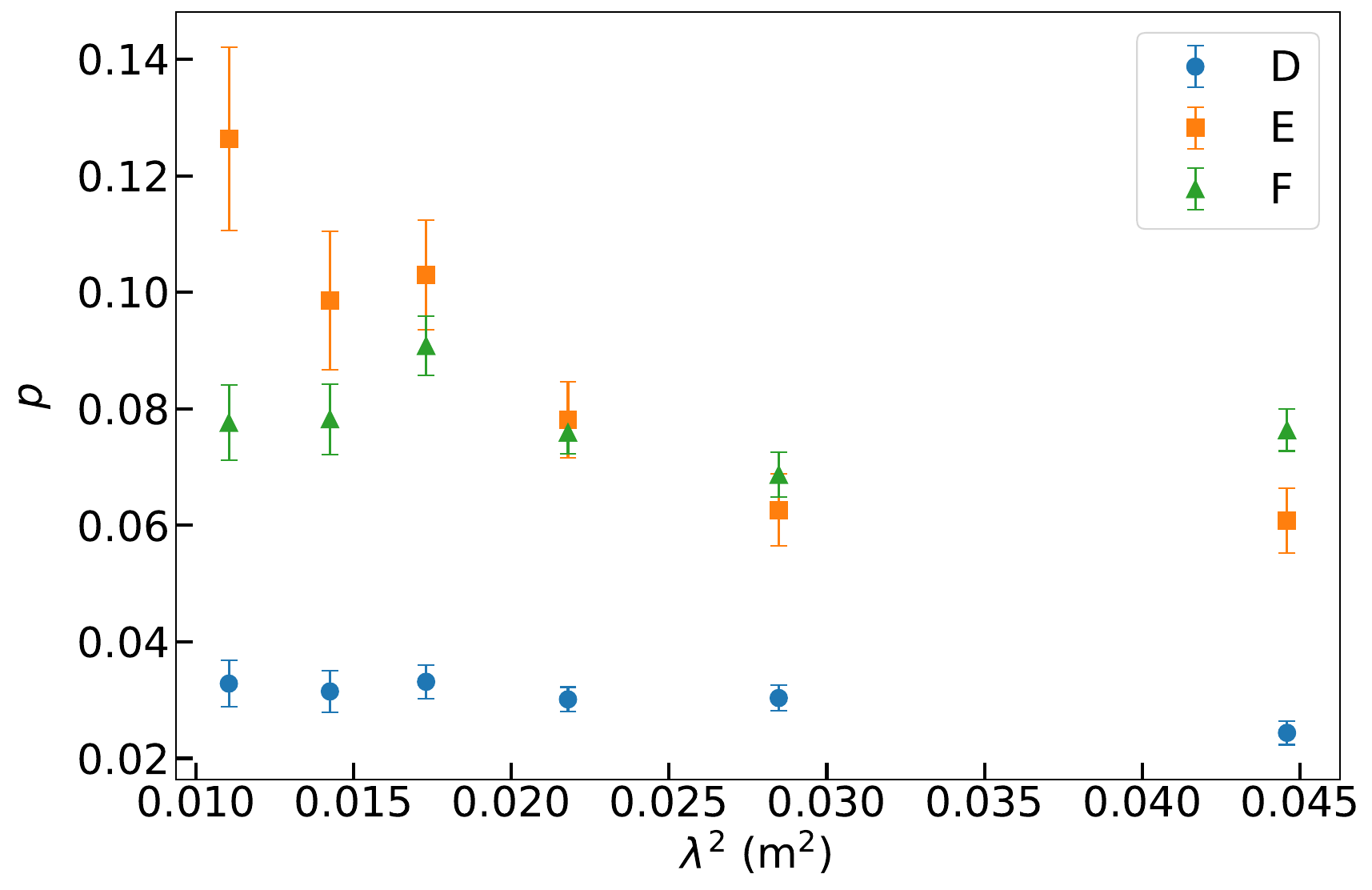}
    \caption{Fractional polarization ($p$) versus wavelength squared ($\lambda^2$) for selected regions of southwest (upper panel) and northeast (lower panel). The regions are outlined in Fig.~\ref{fig:the_degree_of_polarization}.}
    \label{fig:p_lam2}
\end{figure}

\subsubsection{Turbulent magnetic field}
An enhanced turbulent magnetic field is required to accelerate electrons in SNR shocks~\citep[e.g.][]{Bell+04}. In Sect.~\ref{Sect:B_in_SW}, we show that there is a strong turbulent magnetic field in the southwest. Magnetic field fluctuation causes depolarization~\citep{Sokoloff+98}. Therefore, fractional polarization can help constrain the turbulent magnetic field. 

We selected three regions that show different fractional polarizations (Fig.~\ref{fig:the_degree_of_polarization}) from the northeast and southwest, respectively. We averaged $Q$, $U$, and $I$ into six frequency bins from the 80 channel images to increase the signal-to-noise ratio. For each frequency bin, we used the RM of the SNR (Fig.~\ref{fig:RM}) to derotate $Q$ and $U$ to the intrinsic polarization angle before average to account for bandwidth depolarization. We then averaged $Q$, $U$, and $I$ for the six regions A-F and derived $P$, $I$, and $p$. In Fig.~\ref{fig:p_lam2}, we show the resulted $p$ versus $\lambda^2$.  

For the southwest, $p$ does not vary with $\lambda^2$ for the three regions A-C (Fig.~\ref{fig:p_lam2}, upper panel). This indicates that there is little Faraday differential depolarization caused by internal RM. This also agrees with our foreground RM estimate, which results in a small internal RM. The low $p$ is therefore probably caused by the magnetic field fluctuation. According to \citet{Sokoloff+98}, $p_{\rm obs}=p_{\rm i}B_{\perp,\,\rm reg}^2/(B_{\perp,\,\rm reg}^2+\delta B_\perp^2)$, where $B_{\perp,\,\rm reg}$ is the regular transverse magnetic field and $\delta B_\perp$ is the turbulent component. Here, the turbulent magnetic field is assumed to be isotropic. Therefore, the ratio of the turbulent to the regular component of the magnetic field can be estimated as $\delta B_\perp / B_{\perp,\,\rm reg}=\sqrt{p_{\rm i}/p_{\rm obs}-1}$. For $p_{\rm i}=0.7$ and $p_{\rm obs}\approx0.08$ for regions A and C, the ratio is about 3, which agrees with the estimate for the line-of-sight magnetic field estimated in Sect.~\ref{Sect:B_in_SW} taking into account the uncertainties of the foreground RM and the filling factor. For region B, the ratio is even higher. 

For the northeast, $p$ is also nearly constant with $\lambda^2$ for regions D and F, similar to the southwest. The turbulent to regular ratio of the magnetic field of region F is similar to that of regions A and C in the southwest. For region E, $p$ decreases from $\sim0.12$ at $\lambda^2\approx0.012$~m$^2$ to $\sim0.06$ at $\lambda^2\approx0.028$~m$^2$ and keeps roughly the same value until $\lambda^2\approx0.045$~m$^2$. The average internal RM of region E is about 15~rad~m$^{-2}$ after excluding the foreground RM, which cannot account for the factor of 2 decrease of $p$ with Faraday differential depolarization. With an RM scattering of $\sigma_{\rm RM}\approx20$~rad~m$^{-2}$, the beam depolarization~\citep{Sokoloff+98} can produce the $p$ decrease by about 50\% but will result in a much smaller $p$ at the longest $\lambda^2$ of about 0.045~m$^2$. It could be that there are multiple emission components along the line of sight. At shorter wavelengths, all the components can be observed; at the long-wavelength end, only some of the closer components can be observed. The beam depolarization requires the scale of the turbulent magnetic field to be smaller than the beam width, which is about 0.4~pc.     

The ratio of the turbulent to the regular magnetic field is the highest towards region B in the southwest. This is probably due to the interaction between the shock and the interstellar clouds. Observations by \citet{2017+Sano} show that there are HI clumps associated with the southwest of the SNR, and the proton column density of H$_2$ and HI for the southwest is about three times greater than that for the northeast. The interaction between the shock and these clumps results in an amplified magnetic field with huge amplitude fluctuations, and the maximum field strength can be a factor of about 100 larger than the average strength~\citep[][their Figure 5]{2012+Inoue}. This explains the large ratio in the southwest.

However, we also note that the ratios of the turbulent to the regular magnetic field towards regions A and C in the southwest and towards regions E and F in the northeast are very similar. In the northeast, other mechanisms for magnetic field amplification such as the cosmic-ray streaming instability~\citep{Lucek+00} might be more important because the shock velocity is larger. In the southwest, the shock-cloud interaction might be more important because the density of the ambient medium is higher. The turbulent magnetic field is amplified with the regular field. Therefore, the resulted ratios are similar. However, how the magnetic field is amplified is a complicated issue~\citep[e.g.][]{Reynolds+21}, and further investigation is needed.

Young shell-type SNRs are known to accelerate cosmic rays~\citep[e.g.][]{Morlino+17}. There are two scenarios for cosmic-ray electron (CRE) acceleration: quasi-parallel and quasi-perpendicular~\citep[e.g.][]{Jokipii+82, Fulbright+90}. \citet{West+17} showed that a complete turbulent magnetic field can produce an observed radial or tangential field pattern with the selection effects due to the distribution of CREs, and proposed fractional polarization as one of the observables to distinguish the two scenarios. For the southwest, the random field dominates and the radial field pattern shown in Fig.~\ref{fig:Peak polarization} indicates that the quasi-parallel scenario is favored for the mechanism of CRE acceleration there.  

\section{Conclusions}
\label{sec:conclusions}
In this paper, we present a new wideband radio polarization observation of G315.4$-$2.3 using the ATCA, covering the frequency range of 1.1-3.1~GHz. After RFI flagging, the usable frequency range is 1.319-3.023~GHz with some gaps. We obtained $I$, $Q$, and $U$ images at 80 frequency channels, each with approximately 16-MHz bandwidth, by combining the 43 pointings of 6 array configurations with joint deconvolution. We also performed RM synthesis and derived $P$, RM and $p$ images. All images were convolved to a common resolution of $62\arcsec\times33\arcsec$.

We measured the integrated total intensity flux density for the 80 channels. The measurements are consistent with those derived from single-dish observations, indicating that large-scale emission was reserved in our images of G315.4$-$2.3. This allows us to conduct analyses of fractional polarization, which is usually difficult for interferometer observations. 

We obtained a spectral index $\alpha=-0.60\pm0.03$ for the entire SNR by fitting the integrated flux density versus frequency from the measurements in this paper together with previous measurements. The spectral index does not vary much in the SNR with $\alpha=-0.55\pm0.03$ toward the northeast and $\alpha=-0.52\pm0.03$ toward the southwest. 

The foreground RM was estimated to be about 55~rad~m$^{-2}$ based on pulsar RMs and thus the internal RM of most part of the SNR is small and does not cause large depolarization. Based on the internal RM, the regular component of the line-of-sight magnetic field of the southwest was estimated to be about 1.4~$\mu$G, much smaller than the total line-of-sight magnetic field, indicating a large turbulent magnetic field. 

By comparing the fractional polarization of the northeast with that of the southwest, we found that the ratio of the turbulent to the regular component of the magnetic field perpendicular to the line of sight is about 3, confirming the existence of a large turbulent magnetic field. For the southwest, the radial magnetic field pattern indicates that the quasi-parallel scenario is favored for the mechanism of CRE acceleration. For a region in the northeast, there is a large decrease of $p$ with increasing $\lambda^2$, which can be interpreted with beam depolarization, implying that the scale of the turbulent magnetic field is smaller than about 0.4~pc.

In summary, the radio characteristics of the northeast and southwest regions are very similar. They have similar spectra and both have a large turbulent magnetic field. These should be taken into account for future modeling of G315.4$-$2.3. 

\begin{acknowledgements}
We would like to thank the reviewer for the comments that have improved the paper. This research has been supported by the National SKA Program of China (2022SKA0120101). Xin Chen is supported by the Yunnan Provincial Higher Education Institutions Serving Key Industries Technology Project: Doctoral Student Serving Industry Scientific Research Innovation Training Program (Project ID: FWCY-BSPY2024032).

\end{acknowledgements}

\bibliographystyle{aa}
\bibliography{references}
\begin{appendix}
\onecolumn
\section{Faraday spectra}
\begin{figure*}[!htbp]	
    \includegraphics[width=\textwidth]{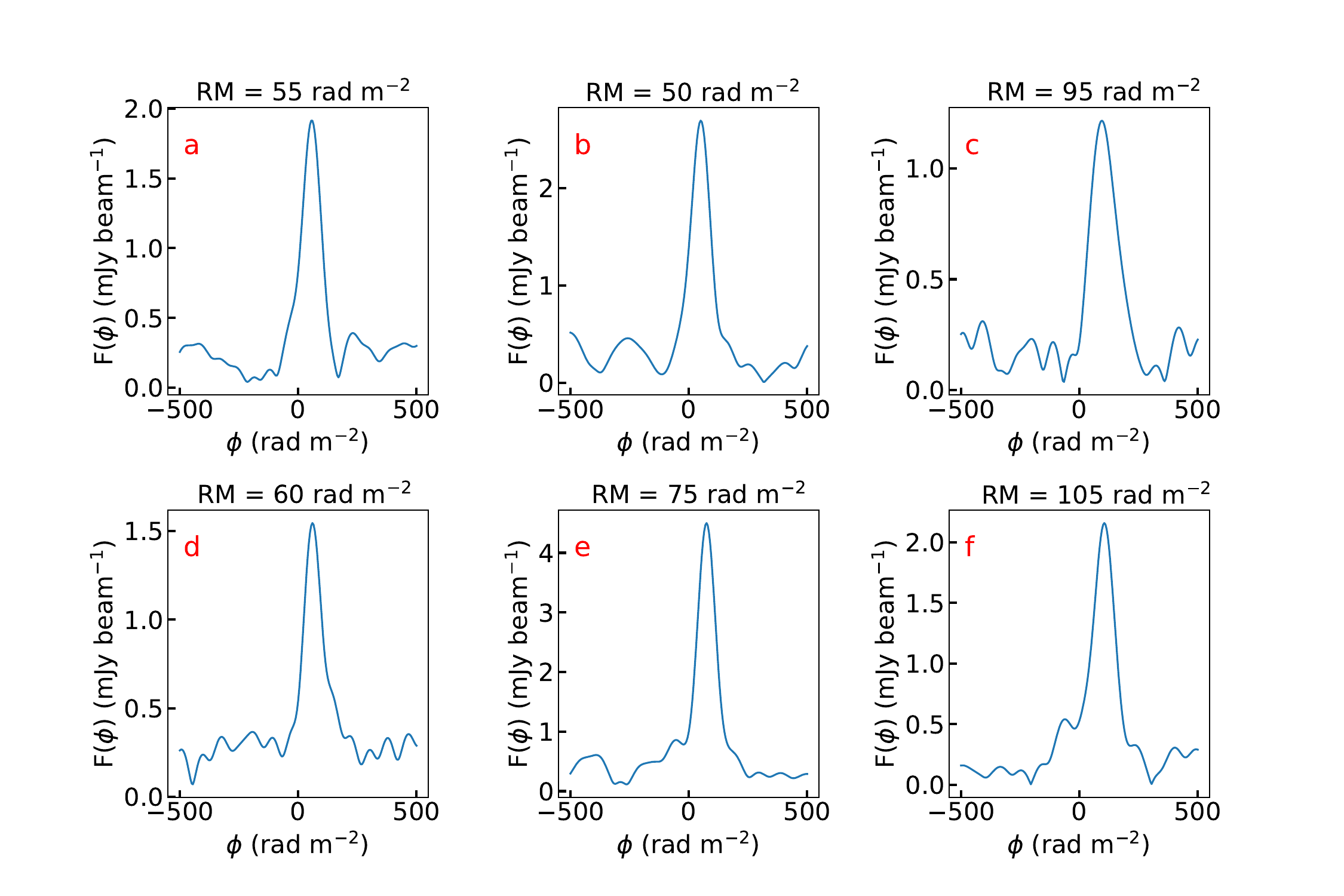}
      \caption{Faraday spectra for pixels a$-$f marked in Fig.~\ref{fig:RM}.}
    \label{fig:check RM}
\end{figure*}
\end{appendix}

\end{document}